# Evidence of decoupling of surface and bulk states in Dirac semimetal Cd$_3$As$_2$

W. Yu[a], D.X. Rademacher[a], N.R. Valdez[a], M.A. Rodriguez[a], T.M. Nenoff[a], and W. Pan[b]

[a] Sandia National Laboratories, Albuquerque, New Mexico 87185, USA

[b] Sandia National Laboratories, Livermore, California 94551, USA

Abstract:

Dirac semimetals have attracted a great deal of current interest due to their potential applications in topological quantum computing, low-energy electronic applications, and single photon detection in the microwave frequency range. Herein are results from analyzing the low magnetic (B) field weak-antilocalization behaviors in a Dirac semimetal Cd$_3$As$_2$ thin flake device. At high temperatures, the phase coherence length $l_\phi$ first increases with decreasing temperature (T) and follows a power law dependence of $l_\phi \propto T^{-0.4}$. Below ~ 3K, $l_\phi$ tends to saturate to a value of ~ 180 nm. Another fitting parameter $\alpha$, which is associated with independence transport channels, displays a logarithmic temperature dependence for T > 3K, but also tends to saturate below ~ 3K. The saturation value, ~ 1.45, is very close to 1.5, indicating three independent electron transport channels, which we interpret as due to decoupling of both the top and bottom surfaces as well as the bulk. This result, to our knowledge, provides first evidence that the surfaces and bulk states can become decoupled in electronic transport in Dirac semimetal Cd$_3$As$_2$.

The discovery of topological materials [1-4] not only enriches our understanding of condensed matter physics, but also is promised to provide exciting new functionalities for future electronic applications. Many forms of topological materials have been predicted. Among them, Dirac semimetals [5,6] have been prominent. In a Dirac semimetal, the conduction and valence bands meet at a single pair of three-dimensional (3D) degenerated Dirac points located at finite $\pm k_D$ momentum points along the [001] direction and display linear dispersion. The gapless crossing, even in the presence of strong spin-orbit interaction in $Cd_3As_2$, is protected by crystal symmetry. Shortly after the prediction of 3D topological Dirac semimetals, materials such as $Cd_3As_2$ were examined, for example, by angle-resolved photoemission spectroscopy (ARPES) [7-9] and scanning tunneling spectroscopy measurements [10] to reveal the Dirac points and linear energy dispersion.

One of the hallmarks of a Dirac semimetal is its unusual topological surface states, the Fermi arc states that connect the projections of the bulk Dirac points on the sample surfaces. Studies of these surface states have generated a great deal of interests in recent years [11-17]; they are a promising candidate for realizing Majorana quasiparticles [18,19] for fault tolerant topological quantum computing [20] and for single photon detection in the microwave frequency range [21]. Recent experiments on dc (direct current) and ac (alternating current) Josephson effects in superconductor-$Cd_3As_2$-superconductor Josephson junctions have shown clear signatures of superconducting topological surface states [11-14]. Moreover, these Josephson effects strongly suggest that there exists a π phase difference between the superconducting surface and bulk states [11]. The exact origin of this π-phase difference is not known at present. As such, it is important to understand electronic transport properties of these surface and bulk states.

The weak antilocalization (WAL) effect has commonly been used to examine the coupling/decoupling of the surface and bulk states in topological materials [22-42]. In a system with spin-momentum locking and the resulting $\pi$ Berry phase, the constructive interference of two time-reversal paths gives rise to a magnetoconductivity correction (Equation 1) [26,34,43] as follows:

$$\Delta\sigma(B) = \sigma_{xx}(B) - \sigma_{xx}(0) = \alpha \times (e^2/\pi h) \times f(B_\phi/B), \qquad (1)$$

where $\sigma_{xx}(B)$ is the conductivity at a magnetic field of B, $\sigma_{xx}(0)$ the conductivity at B = 0, e the electron charge, h the Planck constant, $f(x) \equiv \ln(x) - \Psi(1/2+x)$ with $\Psi$ being the digamma function. $B_\phi$ is the magnetic field associated with the phase coherence length $l_\phi = (\hbar/eB_\phi)^{1/2}$. $\alpha$ is related to the number of independent transport channels. In previous studies on WAL in topological insulators, such as $Bi_2Se_3$, $\alpha$ is determined by coupling/decoupling of the surfaces and bulk [26,28-33,38]. When the surfaces and bulk are strongly coupled, they are treated as one coherent channel and $\alpha = 0.5$. When the top and bottom surfaces are decoupled from the bulk, $\alpha = 1$ due to the existence of two independent channels.

The weak antilocalization effect in Dirac semimetal $Cd_3As_2$ has also been studied by a few groups. The results, particularly the value of $\alpha$, have not been consistent with each other. In two studies on MBE grown $Cd_3As_2$ thin film [35,37], $\alpha$ is considerably smaller than 0.5. Recently, $\alpha = 1$ was obtained in a slightly Sb doped $Cd_3As_2$ sample [42], again grown by the MBE technique. On the other hand, in chemical vapor deposition (CVD) grown $Cd_3As_2$ materials, though weak-antilocalization behaviors have been reported [44,45], a systematic study of $l_\phi$ and $\alpha$ has *not* been carried out.

In this article, we present results from our recent studies of the low-B WAL behaviors at low magnetic fields in a $Cd_3As_2$ thin flake device. The Hikami-Larkin-Nagaoka (HLN) formula [43] is used to fit the WAL behavior to obtain the phase coherence length $l_\phi$ and the constant $\alpha$. It is observed that $l_\phi$ follows a power law dependence with T, $l_\phi \sim T^{-0.4}$, but tends to saturate to ~ 180 nm below T ~ 3K. $\alpha$ displays a logarithmic dependence for T > 3K. It, too, tends to saturate below ~ 3K. Surprisingly, the saturation value, $\alpha \sim 1.45$, is very close to 1.5, indicating three independent transport channels. We argue that these three independent transport channels are due to the decoupling of three, the top and bottom surfaces as well as the bulk, independent channels. This result, to our knowledge, provides first evidence that the surfaces and bulk states can become decoupled in electronic transport in Dirac semimetal $Cd_3As_2$.

$Cd_3As_2$ single crystals are grown in accordance to published methods [46]. Herein, 0.2580 g As powder ($3.4 \times 10^{-3}$ mol, Alfa Aesar, 99.9999% pure) and 1.5501 g Cd powder ($1.38 \times 10^{-2}$ mol, Alfa Aesar, 99.99% pure) were added to a cleaned glass ampoule. The ampoule was cooled and placed under vacuum to $2.0 \times 10^{-5}$ atm, and torch sealed. The ampoule was then placed in a tube furnace and heated at a ramp rate of 20°C/hr to 825°C. It was held at that temperature for 48 hours. It was then cooled at a ramp rate of 3°C/hr to 425°C, and held for 336 hours. At the end of that heating cycle, the ampoule was removed and immediately centrifuged for 3 minutes at 2000 rpm, then quenched in LN2. Black glassy crystals (0.1-1mm) were removed from the ampoule, and confirmed for phase purity by single crystal X-ray diffraction: $Cd_3As_2$, space group = $I4_1/acd$; a = b = 12.6386(4) Å, c = 25.4077(12) Å, V= 4058.5(3) $Å^3$. These lattice parameters are in close agreement with those found in the literature; any discrepancies are minor and may be accounted for within standard errors [46-50]. Figure 1a shows an optical image of an as-grown $Cd_3As_2$ ingot. Fig. 1b shows an XRD powder pattern simulated from the structural parameters refined from single crystal data.

We follow the procedure in Ref. [11] to fabricate our $Cd_3As_2$ thin flake devices for electronic transport measurements. The size of the flake used is approximately 5 μm × 10 μm. E-beam lithography technique is used to define the ohmic contact pads. A metal stack Ti/Au of 10nm/200nm is used to form ohmic contacts to the $Cd_3As_2$ thin flakes. Low temperature electrical measurements are carried out in an Oxford pumped $^3$He system, with a base temperature of ~ 0.3K. By varying the temperatures of its charcoal sorption pump, 1K pot, and a heater close to the sample, the sample temperature can be varied from 0.3K to ~ 50K [51]. A calibrated RuO thermometer, purchased from Scientific Instruments, is used for temperature reading. Four-terminal longitudinal resistance $\rho_{xx}$ and Hall resistance $\rho_{xy}$ are measured using the low-frequency (~ 11 Hz) phase lock-in technique. Two Stanford Research Systems SR830 lock-in amplifiers are used. Lock-in amplifier 1 provides a constant AC voltage (1 V) to induce a current of 10 nA into the sample, through a current-limiting resistor of 100MΩ (much larger than the sample resistance) [52]. This lock-in amplifier measures $\rho_{xx}$. The second lock-in amplifier, synchronized with Lock-in amplifier 1, measures the Hall resistance $\rho_{xy}$. At the excitation current of 10 nA, we estimate that electron self-heating is negligible.

Fig. 2a shows $\rho_{xx}$ as a function of temperature at zero magnetic field. Three regimes with different temperature dependencies are observed. In the temperature range of 10 < T < 50K, it is clearly seen that $\rho_{xx}$ follows a linear T dependence and $\rho_{xx}$ = 928 – 0.77×T, in units of Ω. Between

~ 3K and 10K, $\rho_{xx}$ displays a logarithmic T dependence, as shown in Fig. 2b. This logarithmic temperature dependence is caused by the weak localization effect. Indeed, in a diffusive electron system the destructive quantum interference between two identical self-crossing paths (in which an electron propagates in the opposite directions) leads to an increase in resistivity, which follows a logarithmic temperature dependence [53]. Below 3K, $\rho_{xx}$ increases at a much slower rate and tends to saturate to a value ~ 931 Ω. Fig. 2c shows the magneto-resistivity $\rho_{xx}(B)$, taken at T = 1.3K, in a large magnetic (B) field range. The pronounced weak-antilocalization cusp near the zero magnetic field is clearly seen, consistent with previous work in topological insulators and semimetals [22-42,44,45]. Fluctuations are also observed in magneto-resistivity.

Fig. 3a and 3b show $\rho_{xx}(B)$ and Hall resistivity $\rho_{xy}(B)$, respectively, around B = 0T at a few selected temperatures. As shown in Fig. 3a, the amplitude of fluctuations becomes weaker and eventually disappears at higher temperatures. The Hall resistivity (Fig. 3b) displays a linear B field dependence in the low magnetic fields range around B = 0. All the traces overlap with each other, indicating a constant electron density in the temperature range studied. In Fig. 3c, we plot the area density $n_{2D}$, obtained from the slope of $\rho_{xy}(B)$ as a function of temperature. In the temperature range of 0.5 < T < 38K, $n_{2D}$ ~ $1.5 \times 10^{13}$ cm$^{-2}$. We note that a constant electron density at low temperatures has also been observed before [54]. Moreover, it is believed that a finite electron density in unintentionally doped $Cd_3As_2$ is due to arsenic vacancies [55]. The 3D density is estimated to be $n_{3D}$ ~ $7.5 \times 10^{17}$ cm$^{-3}$, considering the thickness of the thin flake is ~ 200 nm. Consequently, the Fermi energy $E_F$ of the system, estimated by using the following formula (Equation 2), is ~ 100 meV, which is close to the theoretically calculated Lifshitz transition point [48].

$$E_F = \hbar^2(3\pi^2 n_{3D})^{2/3}/2m^*. \quad (2)$$

Here, the effective electron mass in $Cd_3As_2$ is taken as $m^* = 0.03$ (in the units of free electron mass).

We caution here that the Fermi energy is calculated based on a simplified model and does not consider the ellipsoid correction [48]. Moreover, there exists a large discrepancy in the literature on the position of the Lifshitz transition point in $Cd_3As_2$ [6,10,48,56-58]. Theoretically, it was estimated to be ~ 130 meV in Ref. [48]. Experimentally, the measured value differs significantly from one work to another. In Ref. [58], it was estimated ~ 200 meV using the electronic transport technique. On the other hand, a value of as small as ~ 10 meV was estimated in Ref. [10] using the STM technique. Understanding the origin of this discrepancy, though extremely important, is beyond the scope of this work. More future studies are needed.

In the following, we will focus on the weak-antilocalization behavior. As shown in Fig. 3a, the WAL behavior is weakened as T increases. To analyze the weak-antilocalization effect, we follow the previous practices in topological insulators and topological semimetals and use the Hikami-Larkin-Nagaoka (HLN) formula [43] to analyze the weak-antilocalization effect. In fitting the experimentally measured data, we first convert the resistivity to conductivity, $\sigma_{xx}(B) = \rho_{xx}(B)/(\rho_{xx}(B)^2 + \rho_{xy}(B)^2)$. The magneto-conductivity $\Delta\sigma(B) = \sigma_{xx}(B) - \sigma_{xx}(0)$ is then fitted by the formula (Equation 3):

$$\Delta\sigma(B) = \alpha \times (e^2/\pi h) \times [\ln(\hbar/4e l_\phi^2 B) - \Psi(1/2 + \hbar/4e l_\phi^2 B)]. \quad (3)$$

In Fig. 3d, we show a representative fitting at one temperature of 6.8K. A good fitting is seen in the low magnetic field range.

We note here that the HLN formula, developed for two-dimensional electron systems (2DES), fits our data well, considering that our device is 200nm thick. On the other hand, this is not surprising by comparing the $\rho_{xx}(B)$ data in our specimen (Fig. 2c) with the magneto-resistivity data that are well fitted by a 3D WAL model in Ref. [42]. $\rho_{xx}(B)$ in our specimen shows the strong cusp feature typically observed in 2DES. In contrast, the $\rho_{xx}(B)$ data that are well fitted by the 3D WAL model generally shows a quadratic-like B field dependence (see Fig. 3 in Ref. [42]). In fact, the $\rho_{xx}(B)$ curve that shows the cusp feature in their least-doped sample needs to be fitted by the HLN formula, even though the thickness is also about 200 nm (see Supplementary Materials in [42]). In the following, we discuss a couple of possible mechanisms. First, it is known that the spin-orbit coupling in $Cd_3As_2$ is strong. As a result, the spin-orbit scattering time can be significantly shorter than the phase coherence time [42] in our specimen. Consequently, the single coherence channel HLN formula can be valid in the bulk [25]. Second, in our specimen, the phase coherence length is on the order of the device thickness (~ 200nm). This can also make the HLN formula a good approximation in fitting the weak-antilocalization effect for the bulk.

In Figures 4a and 4b, we plot the obtained $l_\phi$ and $\alpha$ as a function of T. $l_\phi$ follows a power-law temperature dependence in the range of ~ 4 to 40 K, $l_\phi \sim T^{-0.4}$, with a power law coefficient of 0.4, suggesting that electron-electron scattering is the main mechanism for the dephasing process [53] in our device. When T is lower than 3K, $l_\phi$ increases at a much slower rate and tends to saturate to a value of ~ 180 nm. The value of $\alpha$ also increases with decreasing T and follows a logarithmic dependence between T ~ 10 and 40K (see Figure 4b). At present, the exact origin of this logarithmic T-depedence is not known. Nevertheless, it indicates that the decoupling of the surfaces and bulk states is a not sudden transition. Rather, it is a gradual process. As T is further reduced below ~ 3K, $\alpha$ also increases at a much slower rate and tends to saturate to a value of ~ 1.45. This low-temperature value of 1.45 is larger than 1, suggesting contributions from more than two independent channels. Considering it is very close to 1.5, we suggest that at low temperatures there exist three independent parallel channels. In a 3D Dirac semimetal like $Cd_3As_2$, these three parallel channels can become possible if both the top and bottom surfaces as well as the bulk all become decoupled. Indeed, both the 2D WAL effect from the top and bottom surfaces and the 3D WAL effect have been observed before [23]. Also, independent surface and bulk channels are observed in the studies of Josephson junctions in $Cd_3As_2$ [11,12]. Mechanisms other than decoupled surfaces and bulk states might also be possible. For example, the Weyl orbits [59] may contribute to the weak anti-localization effect and cause $\alpha$ to be close to 1.5 in Dirac semimetals. More studies are needed.

Our obtained value of $l_\phi$ at low temperatures is consistent with that previously reported in $Cd_3As_2$ samples grown by MBE technique [35,37]. This seems to suggest that $l_\phi$ is independent of how the materials is prepared. On the other hand, the value of $\alpha$ is significantly different. In the MBE grown $Cd_3As_2$ thin films [35,37], $\alpha$ is considerably less than 0.5. It only approaches to 0.5 at low temperatures [37]. This low value of $\alpha$ is probably due to a small film thickness in their samples, which can result in strong coupling of the two surfaces and bulk. Consequently, only one independent transport channel exists. The asymmetric contribution from the surface and the bulk [26] can further reduce $\alpha$ to a value of less than 0.5. Our $Cd_3As_2$ thin flake device is significantly thicker. As a result, at low temperatures, the two (top and bottom) surfaces and the bulk can become decoupled and give rise to three independent channels and, thus, a large $\alpha$ value.

It is surprising that all three parameters, the resistivity, $l_\phi$ and $\alpha$, tend to saturate below ~ 3K. We speculate that the decoupling of top/bottom surfaces and the bulk may be responsible for the tendency. Indeed, if the bulk-surface scattering is the main mechanism for the sample resistivity in $Cd_3As_2$ and dephasing of quantum interference, when all three are decoupled, this scattering mechanism is strongly suppressed. Consequently, $\rho_{xx}$, $l_\phi$, and $\alpha$ can saturate to a constant value, respectively. Additional measurements are ongoing to further explore whether and how the decoupling is related to the $\pi$ phase difference between the surface and bulk states [11].

In conclusion, we have synthesized single crystals of pure $Cd_3As_2$ and fabricated thin flake devices to measure their electronic transport properties. We present results from our systematic studies of the weak-antilocalization effect. The Hikami-Larkin-Nagaoka formula is used to analyze WAL, from which the phase coherence length $l_\phi$ and the constant $\alpha$ (which is related to independence transport channels) are obtained. It is observed that $l_\phi$ follows a power law dependence with T at high temperatures, but saturates to ~ 180 nm below T ~ 3K. $\alpha$ displays a logarithmic dependence for T > 3K, and saturates below ~ 3K. Surprisingly, the saturation value $\alpha$ is very close to 1.5, indicating three independent transport channels probably due to the decoupling of both the top and bottoms surfaces as well as the bulk states in our $Cd_3As_2$ thin flake sample. This observation of decoupled surface channels is expected to have important implications for topologically-protected device applications.

We would like to thank Anna Lima-Sharma and James Park for their help and guidance in the $Cd_3As_2$ crystal growth. The work was supported by a Laboratory Directed Research and

**Figures and Figure Captions:**

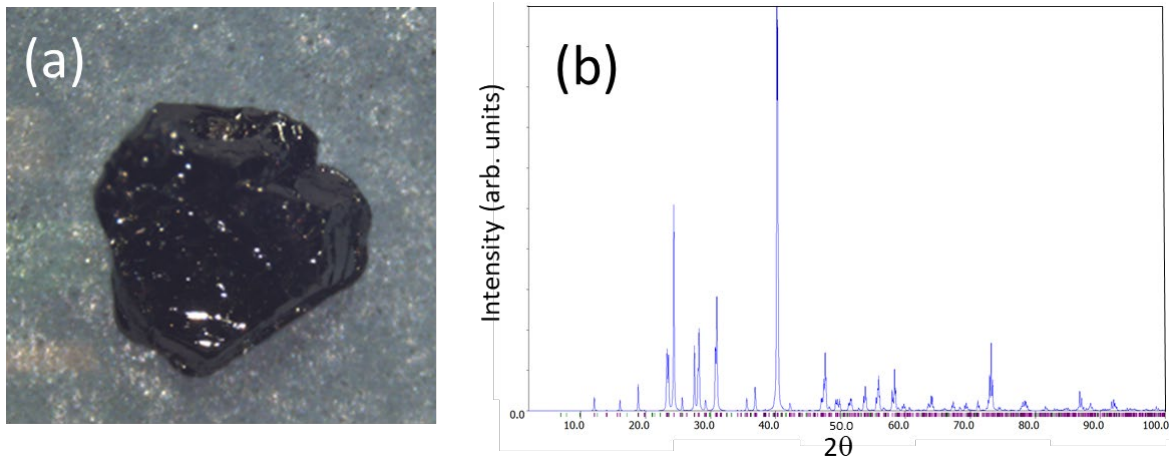

Fig. 1: (a) Optical image of an as-grown Cd$_3$As$_2$ ingot. (b) Powder XRD plot simulated using single crystal data.

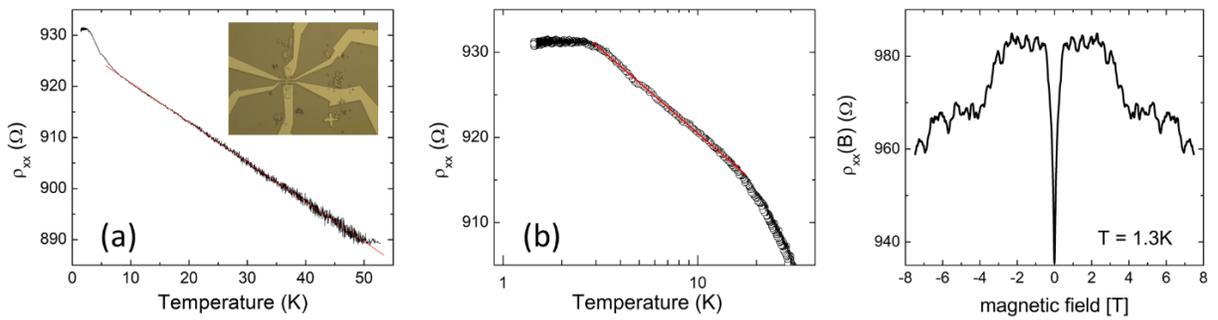

Fig. 2 (a) Resistivity as a function of temperature T. The line is a linear fit. The inset shows a photo of the device studied. (b) Resistivity as a function of T in the logarithmic scale. The line is a linear fit, demonstrating the weak localization behavior. (c) Magneto-resistivity at T = 1.3K. The weak-antilocalization cusp behavior is seen around B = 0T.

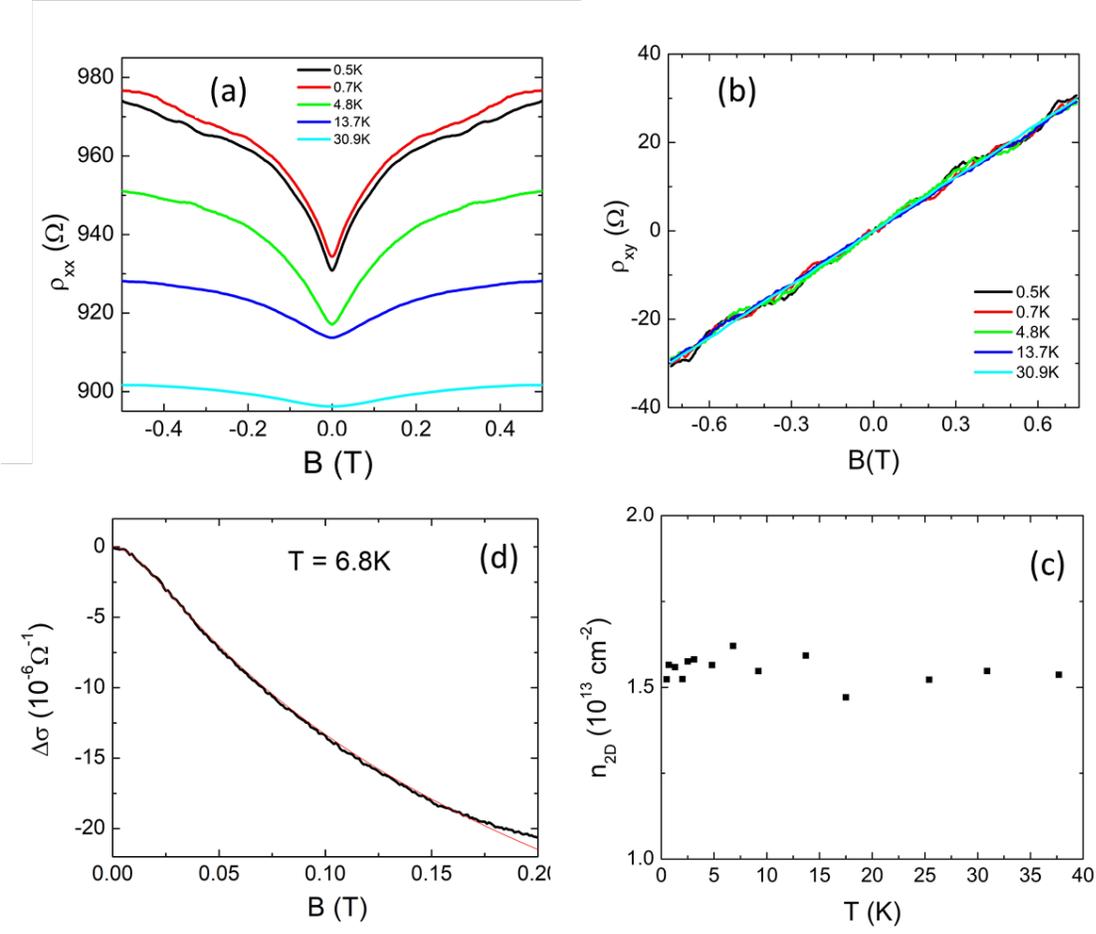

Fig. 3: (a) magneto-resistivity $\rho_{xx}$ around B = 0T at five selected temperatures of 0.5, 0.7, 4.8, 13.7, and 30.9K. (b) Hall resistivity $\rho_{xy}$ as a function of B at the same selected temperature. Linear B dependence is seen. (c) 2D electron density, obtained from the slope of the linear B dependence of $\rho_{xy}$, as a function of temperatures. (d) HLN fitting of the weak antilocalization effect at the temperature of 6.8K.

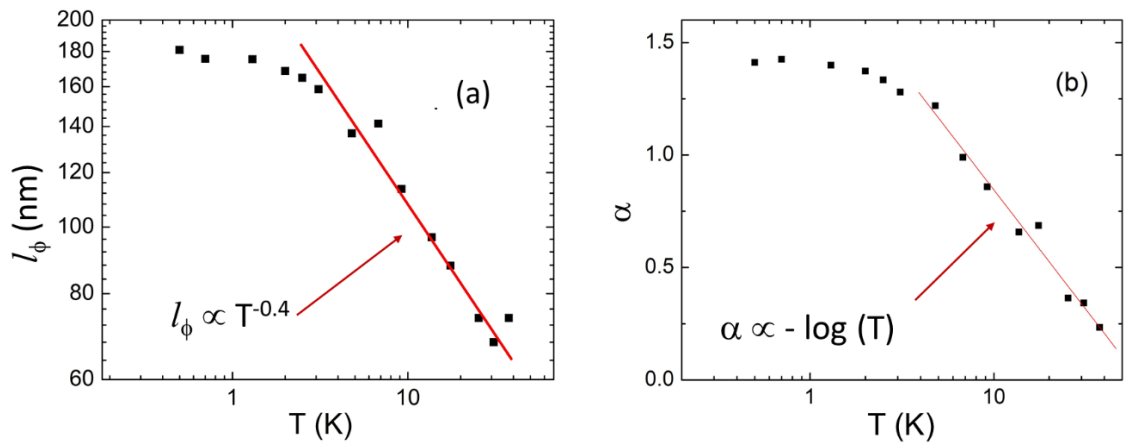

Fig 4: (a) Phase coherence length $l_\phi$ from the HLN fitting as a function of T. A power-law dependence is observed between 5 and 40K. $l_\phi$ saturates to a value of ~ 180 nm below 3K. (b) α as a function of T. It follows a logarithmic T dependence between 5 and 40K but saturates to a value of 1.45 below 3K.